# A simple method for estimating the Lorenz curve

Titithep Sitthiyot [1] & Kanyarat Holasut [2]

Given many popular functional forms for the Lorenz curve do not have a closed-form expression for the Gini index and no study has utilized the observed Gini index to estimate parameter(s) associated with the corresponding parametric functional form, a simple method for estimating the Lorenz curve is introduced. It utilizes three indicators, namely, the Gini index and the income shares of the bottom and the top in order to calculate the values of parameters associated with the specified functional form which has a closed-form expression for the Gini index. No error minimization technique is required in order to estimate the Lorenz curve. The data on the Gini index and the income shares of four countries that have a different level of income inequality, economic, sociological, and regional backgrounds from the United Nations University-World Income Inequality Database are used to illustrate how the simple method works. The overall results indicate that the estimated Lorenz curves fit the actual observations practically well. This simple method could be useful in the situation where the availability of data on income distribution is low. However, if more data on income distribution are available, this study shows that the specified functional form could be used to directly estimate the Lorenz curve. Moreover, the estimated values of the Gini index calculated based on the specified functional form are virtually identical to their actual observations.

[1] Chulalongkorn University, Bangkok, Thailand. [2] Khon Kaen University, Khon Kaen, Thailand. ✉email: titithep@cbs.chula.ac.th

## Introduction

The Lorenz curve was devised by an American economist named Max O. Lorenz in 1905 as a method for measuring the concentration of wealth. It represents a graphical relationship between the cumulative normalized rank of the population from the poorest to the richest and the cumulative normalized wealth held by these population from the poorest to the richest. For more than a century, the Lorenz curve has been widely used for illustrating the distributions of income and wealth and for examining inequality in such distributions. Jordá et al. (2021) note that income inequality would be relatively simple to estimate if individual records on personal or household income data were available. Unfortunately, much of the existing research on economic inequality has been plagued by a lack of individual data. However, periodic report of certain summary statistics on the income distribution has become quite common. The United Nations University-World Income Inequality Database (UNU-WIID), the World Bank's PovcalNet, and the World Income Database (WID) are the largest cross-country databases that provide grouped income data, typically including the information on income and population shares. This type of grouped data depicts sparse points of the Lorenz curve, which makes defining a method to connect those points an essential requisite for estimating inequality measures.

By using the grouped data on income, the Lorenz curve could be estimated (1) by employing an interpolation technique, (2) by assuming a statistical distribution of income and deriving the corresponding Lorenz curve, and (3) by specifying a parametric functional form for the Lorenz curve (Paul and Shankar, 2020). The interpolation technique assumes the homogeneity of income within sub-groups and thus underestimates inequality in income distribution (Paul and Shankar, 2020) whereas no single statistical distribution function has proved to fit the entire income distribution. According to Chakrabarti et al. (2013), there tends to be an agreement among scholars in economics, statistics, and physics that the upper tail of income distribution can be well described by a power-law distribution. For the lower part of the distribution, it is still debatable whether income follows gamma or log-normal distribution. If the choice is not a valid candidate for representing the income distribution, the estimates on the income shares and inequality measures might be severely affected by misspecification bias (Jordá et al., 2021).

Given the limitations of the interpolation technique and no single statistical distribution that fits the entire income distribution, numerous studies have suggested a variety of parametric functional forms to directly estimate the Lorenz curve. Examples include Kakwani and Podder (1973, 1976), Kakwani (1980), Rasche et al. (1980), Aggarwal (1984), Gupta (1984), Arnold (1986), Rao and Tam (1987), Villaseñor and Arnold (1989), Basmann et al. (1990), Ortega et al. (1991), Chotikapanich (1993), Ogwang and Rao (1996, 2000), Ryu and Slottje (1996), Sarabia (1997), Sarabia et al. (1999, 2001, 2010, 2015, 2017), Sarabia and Pascual (2002), Rohde (2009), Helene (2010), Wang and Smyth (2015), Fellman (2018), Tanak et al. (2018), Paul and Shankar (2020), and Sitthiyot et al. (2020). However, many existing widely used functional forms often cited in the literature do not have a closed-form expression for the Gini index, making it computationally inconvenient to calculate since they require the valuation of the beta function, for example, Kakwani and Podder (1976), Kakwani (1980), Rasche et al. (1980), and Ortega et al. (1991), or the confluent hypergeometric function, for example, Rao and Tam (1987). For other popular functional forms often cited in the literature that have an explicit mathematical solution for the Gini index, for example, Kakwani and Podder (1973), Aggarwal (1984), Gupta (1984), Chotikapanich (1993), and Rohde (2009) the step from using the estimated parameter(s) associated with the specified functional forms to calculate the respective Gini index is relatively less complicated than the step starting from the observed Gini index and working backward to compute the value(s) of the parameter(s) associated with the corresponding specified functional forms. To our knowledge, no study has attempted to conduct a reverse-engineer by utilizing the observed Gini index in order to calculate the value(s) of the parameter(s) associated with the specified functional form despite the availability of data on the Gini index periodically published by countries and/or international organizations.

To address the key issues with regard to the existing parametric functional forms as discussed above, this study introduces a method for estimating the Lorenz curve. Our method is simple and straightforward. It utilizes three indicators, namely, the Gini index, the income share of the bottom, and that of the top in order to calculate the values of parameters associated with the specified functional form that is based on the weighted average of the exponential function and the functional form implied by Pareto distribution without relying on any complicated error minimization technique. Note that the Gini index provides the information about the overall shape of the Lorenz curve especially at the center while the data on income shares of the bottom and the top give the additional information at the tails of the income distribution.

To demonstrate the simple method, we use the data on the Gini index and the income shares of four countries that differ in the degree of income inequality, economic, sociological, and regional backgrounds from the UNU-WIID. They are Malta, Taiwan, the United States of America (USA), and Cote d'Ivoire. This simple method could be useful in the situation where the availability of data on income distribution is low since it does not require a collection of data at the micro-level. If more data on the income distribution are available, the specified parametric functional form that is based on the weighted average of the exponential function and the functional form implied by Pareto distribution could be used to directly estimate the Lorenz curve. In addition, the Gini index can be conveniently computed since the specified functional form has a closed-form expression for the Gini index. Furthermore, we compare the performance of our specified functional form to that of Kakwani (1980) which, according to Cheong (2002), has the overall superior performance to other functional forms used for estimating the Lorenz curve. Tanak et al. (2018) also note that the functional form proposed by Kakwani (1980) is considered as the best performer among a number of different functional forms for the Lorenz curve in fitting to the data. Given the use of the Lorenz curve and the Gini index in various contexts besides income,[1] our simple method and the alternative functional form for estimating the Lorenz curve could be useful for broad scientific disciplines that employ the Lorenz curve and the Gini index for analyzing size distributions of non-negative quantities and inequality.

## Methods

Let $x$ denote the cumulative normalized rank of income, $y$ denote the cumulative normalized income, $k$ and $P$ be parameters. While there is a vast family of existing and already known functional forms for the Lorenz curve that could be used in combination by assigning a weight between 0 and 1 to each functional form such that all weights are added up to 1,[2] the specified functional form considered as suitable for developing our simple method is based on the weighted average of two well-known functional forms which are the exponential function ($y(x) = x^P$) and the functional form implied by Pareto distribution ($y(x) = 1-(1-x)^{\frac{1}{P}}$). Note that, if we separately integrate the exponential function and the functional form implied by Pareto distribution from 0 to 1, it can be seen that both functional forms have the same area under the Lorenz curve which is equal to $\frac{1}{P+1}$. By assigning the weight $(1-k)$ to the exponential function and the weight $k$ to the functional form implied by Pareto distribution, the specified functional form based on the weighted average of these two well-known functional forms is as follows:

$$y(x) = (1-k) * x^P + k * \left(1 - (1-x)^{\frac{1}{P}}\right) \tag{1}$$

$$0 \le k \le 1$$

$$1 \le P$$

This specified functional form satisfies all necessary and sufficient conditions for the Lorenz curve which are: $y(0) = 0$, $y(1) = 1$, $y(x)$ is convex, $\frac{dy}{dx} \ge 0$, and $\frac{d^2y}{dx^2} \ge 0$. The condition $\frac{d^2y}{dx^2} \ge 0$ also considers the case where everyone has the same income. While different scientific disciplines may have their own theoretical justifications when using the specified functional form for the Lorenz curve as a tool to investigate size distributions of non-negative quantities and calculate statistical evenness measures, from an economic point of view, the parameter $P$ represents the degree of inequality in income distribution as measured by the Gini index. The parameter $k$ is the weight that controls the curvature of the Lorenz curve such that the Gini index remains unchanged since, for a given value of parameter $P$, there are infinite values of parameter $k$ that could give an identical value of the Gini index. The parameter $k$ thus provides the key information about countries' income shares in case their Lorenz curves intersect. In addition, from the policy standpoint, the key advantage of specifying the functional form that is based on the weighted average of the exponential function and the functional form implied by Pareto distribution, besides its simplicity, is that the shape of the estimated Lorenz curve could be handily adjusted through the change in parameter $k$ while the value of the Gini index is kept constant which may not be easily done for other linear convex combinations of functional forms for the Lorenz curve. Note however that even though our specified functional form is constructed based on the existing and well-known functional forms for the Lorenz curve, to our knowledge, no study has employed a parametric functional form for approximating the Lorenz curve by combining the exponential function and the functional form implied by Pareto distribution in this way before. The one that comes close is the parametric functional form based on a linear convex combination of the egalitarian line, the power Lorenz curve, and the classical Pareto Lorenz curve proposed by Sarabia (1997). Based on our specified functional form as shown in Eq. (1), the area under the Lorenz curve and the explicit mathematical solution for the Gini index ($Gini$) can be conveniently calculated as shown in Eqs. (2) and (3), respectively.

$$\int_0^1 y(x)dx = \frac{1}{P+1} \tag{2}$$

$$Gini = 1 - 2 * \int_0^1 y(x)dx = \frac{P-1}{P+1} \tag{3}$$

$$0 \le Gini \le 1$$

The Gini index takes the values between 0 and 1. The closer the index is to 0, the more equal the distribution of income while the closer the index is to 1, the more unequal the income distribution as illustrated in Fig. 1.

By utilizing the available data on the observed Gini index as periodically published by countries and/or international organizations, we can compute the value of parameter $P$ as follows:

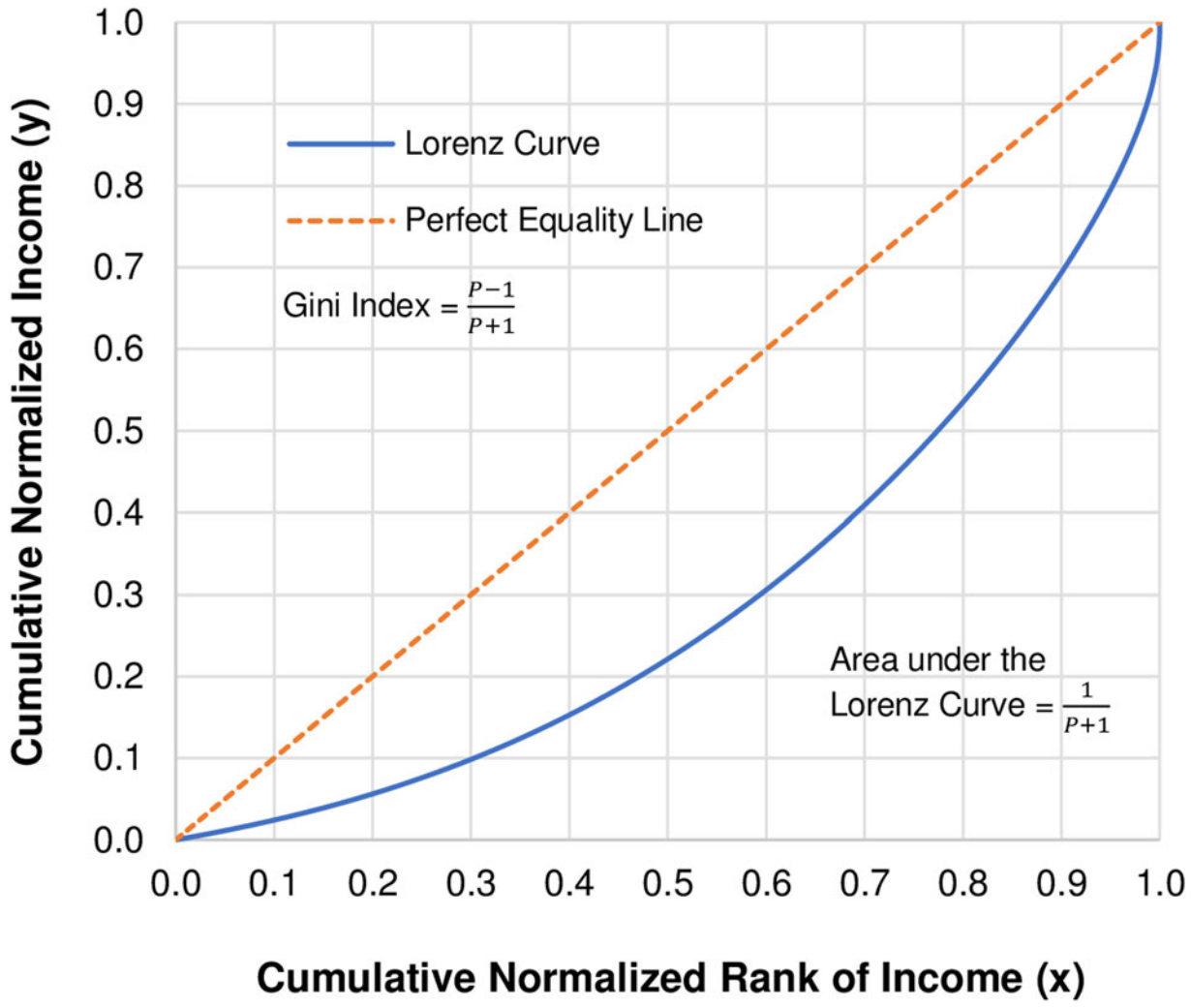

**Fig. 1 The Lorenz curve.** Given the specified functional form that is constructed based on the weighted average of the exponential function and the functional form implied by Pareto distribution, the closed-form expression for the Gini index can be conveniently computed as $\frac{P-1}{P+1}$.

$$P = \frac{1 + Gini}{1 - Gini} \tag{4}$$

To calculate the value of parameter $k$, we use the value of parameter $P$ obtained from Eq. (4) and utilize the available data on the income shares of the bottom and the top as published by countries and/or international organizations. Let $B_m$ be the income share of the bottom $m$% and $T_m$ be the income share of the top $m$%. Based on the specified functional form for the Lorenz curve as shown in Eq. (1), the ratio of income shares of the bottom $m$% to that of the top $m$% could be expressed as follows:

$$\frac{B_m}{T_m} = \frac{(1-k) * m^P + k * \left[1 - (1-m)^{\frac{1}{P}}\right]}{1 - \left((1-k) * n^P + k * \left[1 - (1-n)^{\frac{1}{P}}\right]\right)} \tag{5}$$

$$n = 1 - m$$

$$0 \le m \le 1$$

Next, let $R_m = \frac{B_m}{T_m}$, $a = m^P$, $b = (1-m)^{\frac{1}{P}}$, $c = n^P$, and $d = (1-n)^{\frac{1}{P}}$. Equation (5) could be rewritten as follows:

$$R_m = \frac{(1-k) * a + k * (1-b)}{1 - ((1-k) * c + k * (1-d))} \tag{6}$$

Given the values of $R_m$, $a$, $b$, $c$, and $d$, the value of parameter $k$ can be computed as shown in Eq. (7).

$$k = \frac{a - R_m + c * R_m}{c * R_m - R_m + d * R_m + a + b - 1} \tag{7}$$

After obtaining the values of parameters $P$ and $k$, the entire Lorenz curve can be retrieved. This Lorenz curve is estimated by using only three indicators which are the Gini index, the income share of the bottom $m$%, and that of the top $m$% as published by countries and/or international organizations.[3] Note that no complicated error minimization technique is required in order to estimate the Lorenz curve. Thus, there is no need to resort to writing algorithms for estimating parameters using a software package such as R as provided in Bourguignon et al. (2016) and Jordá et al. (2021).

**Table 1 The estimated parameters *P* and *k* for the Lorenz curves of four countries calculated using the simple method that utilizes the data on the Gini index and the income shares of the bottom 10% and the top 10% as published in the UNU-WIID.**

| Year | Country | Gini Index | $B_{10}$ | $T_{10}$ | $B_{10}/T_{10}$ | $P_{simple\ method}$ | $k_{simple\ method}$ | $R^2$ |
|---|---|---|---|---|---|---|---|---|
| 2018 | Malta | 0.287 | 0.036 | 0.230 | 0.157 | 1.81 | 0.47 | 0.9970 |
| 2016 | Taiwan | 0.315 | 0.034 | 0.252 | 0.133 | 1.92 | 0.49 | 0.9929 |
| 2016 | USA | 0.411 | 0.018 | 0.305 | 0.059 | 2.40 | 0.31 | 0.9713 |
| 2015 | Côte d'Ivoire | 0.590 | 0.007 | 0.477 | 0.016 | 3.88 | 0.22 | 0.9095 |

To demonstrate our simple method for approximating the Lorenz curve, we employ the latest data on the Gini index and the income shares by decile and by percentile of four countries, namely, Malta, Taiwan, USA, and Côte d'Ivoire between 2015 and 2018 from the UNU-WIID. The main reason that we use the UNU-WIID is that it is the only database that has the data on the Gini index and the income shares by decile and by percentile, all of which, according to Jordá et al. (2021), are computed from the same underlying microdata. Note that these countries are mainly chosen based on their differences in the degree of income inequality as measured by the Gini index and also to reflect economic, sociological, and regional diversities. As suggested by Dagum (1977), a good parametric functional form for the Lorenz curve has to be able to describe income distributions of different countries, regions, socioeconomic groups, and in different time periods. It should also provide a good fit of the entire range of income distribution since all observations are relevant for accurate measurement of income inequality, supporting social and income policies, as well as determining taxation structure.

Following Chotikapanich (1993), Cheong (2002), Tanak et al. (2018), and Paul and Shankar (2020), five goodness-of-fit statistics, namely, coefficient of determination ($R^2$), mean square error (MSE), mean absolute error (MAE), maximum absolute error (MAS), and information inaccuracy measure (IIM) are used to gauge how close the estimated income shares by decile are to the empirical observations. Moreover, we conduct the Kolmogorov-Smirnov test (K-S test) to compare whether the estimated income shares by decile are statistically different from the actual observations with the null hypothesis being no difference between the two. According to Sitthiyot et al. (2020), the K-S test is commonly used to determine if two datasets differ significantly. Its advantage is that it makes no assumption about the distribution of data.

In addition to the simple method which could be applied when the availability of the data on income distribution is low, we employ the curve fitting technique based on minimizing the error sum of squares in order to demonstrate that if more data on the income distribution are available, our specified parametric functional form that is constructed based on the weighted average of the exponential function and the functional form implied by Pareto distribution could be used to directly estimate the Lorenz curve. The fitted Lorenz curve based on the error sum of square minimization is then used to calculate the Gini index whose values would be compared to the actual observations. Moreover, we compare the performance of our specified functional form to that of Kakwani (1980) which is considered as the best performer among existing widely used functional forms for the Lorenz curve in fitting to the actual observations according to Cheong (2002) and Tanak et al. (2018). In this study, the Microsoft Excel Solver program, which is available in most, if not all, computers, is used for estimating the parameters and calculating the estimated value of the Gini index. As noted by Dagum (1977), from a viewpoint of computational cost and the acceptance of the specified functional form for practical purposes, a simple method of parameter estimation is always an advantage.

## Results and discussion

**The simple method using the Gini index and the income shares of the bottom 10% and the top 10%**. Table 1 reports the estimated parameters $P$ and $k$ for the Lorenz curves of four countries, namely, Malta, Taiwan, USA, and Côte d'Ivoire using the simple method that utilizes the data on the Gini index, the income shares of the bottom 10% and the top 10% as published in the UNU-WIID. The results as shown in Table 1 indicate that all four estimated Lorenz curves fit the empirical observations practically well with the values of $R^2$ ranging between 0.9095 and 0.9970. The results from $R^2$ suggest that the lower the level of income inequality as measured by the Gini index, the more fitted the estimated Lorenz curve is to the empirical observations.

We then use the fitted Lorenz curves to calculate the values of income shares by decile of four countries and compare them to their actual observations. The actual and the estimated income shares by decile of the four countries as well as the values of goodness-of-fit statistics, namely, MSE, MAE, MAS, and IIM are reported in Table 2. All values of goodness-of-fit measures suggest that the estimated income shares by decile of each country do not differ significantly from the actual observations. In addition to the goodness-of-fit statistics, we perform the K-S test in order to compare whether the estimated income shares by decile are statistically different from their actual observations. The results from the K-S test as shown in Table 2 indicate that the differences between the estimated income shares by decile and the actual observations are statistically not significant with $p$-values ranging between 0.975 and 1.000. The results shown in Table 2 also suggest that the lower the degree of country's income inequality as measured by the Gini index, the closer the estimated income shares by decile are to the actual observations.

**The simple method using the Gini index and the income shares of the bottom 5% and the top 5%**. Next, we employ the simple method to estimate the parameters $P$ and $k$ associated with the Lorenz curve by using the data on the Gini index, the income share of the bottom 5%, and that of the top 5% of four countries as published in the UNU-WIID. The results are reported in Table 3. The overall results indicate that the estimated Lorenz curves fit the actual observations reasonably well with the values of $R^2$ ranging between 0.8520 and 0.9980. Similar to the results shown in Table 1, the values of $R^2$ suggest that the lower the degree of income inequality as measured by the Gini index, the better fitted the estimated Lorenz curve is to the actual observations.

The fitted Lorenz curves approximated by using the simple method that utilizes the data on the Gini index, the income share of the bottom 5%, and that of the top 5% are then used to calculate the income shares by decile of four countries in the sample. The results are reported in Table 4. All values of goodness-of-fit statistics which are MSE, MAE, MAS, and IIM suggest that the estimated income shares by decile do not significantly differ from the actual observations. The K-S test confirms that the differences between the two are statistically not significant with $p$-values ranging between 0.975 and 1.000. Similar to the results shown in Table 2, the results shown in Table 4 suggest that the lower the level of the country's income inequality as measured by the Gini index, the closer the estimated decile income shares are to the actual observations.

**Table 2 The comparison between the actual and the estimated income shares by decile calculated using the simple method that utilizes the data on the Gini index and the income shares of the bottom 10% and the top 10%.**

| Decile | Malta | | Taiwan | | USA | | Côte d'Ivoire | |
|---|---|---|---|---|---|---|---|---|
| | Actual | Estimate | Actual | Estimate | Actual | Estimate | Actual | Estimate |
| D1 | 0.0360 | 0.0349 | 0.0336 | 0.0321 | 0.0179 | 0.0160 | 0.0074 | 0.0059 |
| D2 | 0.0510 | 0.0487 | 0.0491 | 0.0446 | 0.0344 | 0.0260 | 0.0178 | 0.0078 |
| D3 | 0.0620 | 0.0609 | 0.0590 | 0.0566 | 0.0457 | 0.0392 | 0.0263 | 0.0128 |
| D4 | 0.0720 | 0.0727 | 0.0684 | 0.0687 | 0.0572 | 0.0549 | 0.0353 | 0.0228 |
| D5 | 0.0830 | 0.0843 | 0.0779 | 0.0811 | 0.0693 | 0.0727 | 0.0459 | 0.0396 |
| D6 | 0.0940 | 0.0964 | 0.0890 | 0.0941 | 0.0832 | 0.0926 | 0.0583 | 0.0649 |
| D7 | 0.1070 | 0.1094 | 0.1022 | 0.1084 | 0.1005 | 0.1148 | 0.0759 | 0.1006 |
| D8 | 0.1210 | 0.1245 | 0.1199 | 0.1252 | 0.1245 | 0.1401 | 0.1026 | 0.1489 |
| D9 | 0.1440 | 0.1454 | 0.1493 | 0.1487 | 0.1625 | 0.1718 | 0.1535 | 0.2142 |
| D10 | 0.2300 | 0.2229 | 0.2517 | 0.2405 | 0.3046 | 0.2719 | 0.4769 | 0.3826 |
| MSE | 0.00001 | | 0.00003 | | 0.00018 | | 0.00159 | |
| MAE | 0.0023 | | 0.0040 | | 0.0104 | | 0.0276 | |
| MAS | 0.0071 | | 0.0112 | | 0.0327 | | 0.0943 | |
| IIM | 0.0003 | | 0.0012 | | 0.0063 | | 0.0458 | |
| K-S test | *D*-statistic = 0.1000 | | *D*-statistic = 0.1000 | | *D*-statistic = 0.1000 | | *D*-statistic = 0.200 | |
| | *p*-value = 1.000 | | *p*-value = 1.000 | | *p*-value = 1.000 | | *p*-value = 0.975 | |

**Table 3 The estimated parameters *P* and *k* for the Lorenz curves of four countries calculated using the simple method that utilizes the data on the Gini index and the income shares of the bottom 5% and the top 5% as published in the UNU-WIID.**

| Year | Country | Gini index | $B_5$ | $T_5$ | $B_5/T_5$ | $P_{simple\ method}$ | $k_{simple\ method}$ | $R^2$ |
|---|---|---|---|---|---|---|---|---|
| 2018 | Malta | 0.287 | 0.016 | 0.139 | 0.115 | 1.81 | 0.48 | 0.9980 |
| 2016 | Taiwan | 0.315 | 0.014 | 0.156 | 0.088 | 1.92 | 0.39 | 0.9759 |
| 2016 | USA | 0.411 | 0.006 | 0.196 | 0.029 | 2.40 | 0.17 | 0.9223 |
| 2015 | Côte d'Ivoire | 0.590 | 0.002 | 0.350 | 0.006 | 3.88 | 0.09 | 0.8520 |

**Table 4 The comparison between the actual and the estimated income shares by decile calculated using the simple method that utilizes the data on the Gini index and the income shares of the bottom 5% and the top 5%.**

| Decile | Malta | | Taiwan | | USA | | Côte d'Ivoire | |
|---|---|---|---|---|---|---|---|---|
| | Actual | Estimate | Actual | Estimate | Actual | Estimate | Actual | Estimate |
| D1 | 0.0360 | 0.0354 | 0.0336 | 0.0283 | 0.0179 | 0.0106 | 0.0074 | 0.0026 |
| D2 | 0.0510 | 0.0490 | 0.0491 | 0.0425 | 0.0344 | 0.0220 | 0.0178 | 0.0044 |
| D3 | 0.0620 | 0.0610 | 0.0590 | 0.0561 | 0.0457 | 0.0372 | 0.0263 | 0.0097 |
| D4 | 0.0720 | 0.0725 | 0.0684 | 0.0695 | 0.0572 | 0.0552 | 0.0353 | 0.0208 |
| D5 | 0.0830 | 0.0840 | 0.0779 | 0.0832 | 0.0693 | 0.0754 | 0.0459 | 0.0395 |
| D6 | 0.0940 | 0.0959 | 0.0890 | 0.0973 | 0.0832 | 0.0978 | 0.0583 | 0.0677 |
| D7 | 0.1070 | 0.1088 | 0.1022 | 0.1124 | 0.1005 | 0.1222 | 0.0759 | 0.1076 |
| D8 | 0.1210 | 0.1239 | 0.1199 | 0.1294 | 0.1245 | 0.1492 | 0.1026 | 0.1610 |
| D9 | 0.1440 | 0.1450 | 0.1493 | 0.1519 | 0.1625 | 0.1804 | 0.1535 | 0.2310 |
| D10 | 0.2300 | 0.2243 | 0.2517 | 0.2295 | 0.3046 | 0.2500 | 0.4769 | 0.3557 |
| MSE | 0.00001 | | 0.00009 | | 0.00049 | | 0.00259 | |
| MAE | 0.0019 | | 0.0074 | | 0.0170 | | 0.0354 | |
| MAS | 0.0057 | | 0.0222 | | 0.0546 | | 0.1212 | |
| IIM | 0.0002 | | 0.0036 | | 0.0179 | | 0.0800 | |
| K-S test | *D*-statistic = 0.100 | | *D*-statistic = 0.100 | | *D*-statistic = 0.100 | | *D*-statistic = 0.200 | |
| | *p*-value = 1.000 | | *p*-value = 1.000 | | *p*-value = 1.000 | | *p*-value = 0.975 | |

Based on the results from using the data on income shares of the bottom 10% and the top 10% vs. using the data on income shares of the bottom 5% and the top 5% to estimate the decile income shares, it can be concluded that the performance of our simple method, which utilizes three indicators, namely, the Gini index and the income shares of the bottom and the top, depends on the level of income inequality as measured by the Gini index and the income shares at the tails of the income distribution. Even though all values of goodness-of-fit measures as shown in Table 4 indicate that the estimated income shares by decile do not statistically differ from the actual observations, the results in Table 2 indicate that our simple method tends to perform slightly better when using the data on income shares of the bottom 10% and the top 10% relative to the data on income shares of the bottom 5% and the top 5%. Given these results, it should be noted that, in the circumstance where the only available data are the income shares at the extreme tails, for example, the bottom 5% and the top 5% and the level of income inequality as measured by the Gini index is high, for example, 0.590 as for the case of Côte d'Ivoire, our simple method would give a cruder approximation for the Lorenz curve.

**Table 5 The estimated parameters $P$ and $k$ for the Lorenz curves of four countries calculated using the error minimization method that utilizes all available grouped data on decile income shares as published in the UNU-WIID.**

| Year | Country | Gini Index | $B_{10}$ | $T_{10}$ | $B_{10}/T_{10}$ | $P_{\text{error minimization method}}$ | $k_{\text{error minimization method}}$ | $R^2$ |
|---|---|---|---|---|---|---|---|---|
| 2018 | Malta | 0.287 | 0.036 | 0.230 | 0.157 | 1.81 | 0.53 | 1.0000 |
| 2016 | Taiwan | 0.315 | 0.034 | 0.252 | 0.133 | 1.92 | 0.60 | 1.0000 |
| 2016 | USA | 0.411 | 0.018 | 0.305 | 0.059 | 2.40 | 0.52 | 0.9999 |
| 2015 | Côte d'Ivoire | 0.590 | 0.007 | 0.477 | 0.016 | 3.86 | 0.64 | 0.9995 |

**Table 6 The comparison between the actual and the estimated income shares by decile as well as the comparison between the observed and the estimated Gini index calculated using the error minimization method that utilizes all available grouped data on decile income shares as published in the UNU-WIID.**

| Decile | Malta | | Taiwan | | USA | | Côte d'Ivoire | |
|---|---|---|---|---|---|---|---|---|
| | Actual | Estimate | Actual | Estimate | Actual | Estimate | Actual | Estimate |
| D1 | 0.0360 | 0.0376 | 0.0336 | 0.0366 | 0.0179 | 0.0242 | 0.0074 | 0.0172 |
| D2 | 0.0510 | 0.0500 | 0.0491 | 0.0470 | 0.0344 | 0.0320 | 0.0178 | 0.0193 |
| D3 | 0.0620 | 0.0612 | 0.0590 | 0.0572 | 0.0457 | 0.0423 | 0.0263 | 0.0232 |
| D4 | 0.0720 | 0.0720 | 0.0684 | 0.0676 | 0.0572 | 0.0545 | 0.0353 | 0.0298 |
| D5 | 0.0830 | 0.0829 | 0.0779 | 0.0785 | 0.0693 | 0.0686 | 0.0459 | 0.0401 |
| D6 | 0.0940 | 0.0943 | 0.0890 | 0.0903 | 0.0832 | 0.0847 | 0.0583 | 0.0554 |
| D7 | 0.1070 | 0.1068 | 0.1022 | 0.1037 | 0.1005 | 0.1033 | 0.0759 | 0.0771 |
| D8 | 0.1210 | 0.1219 | 0.1199 | 0.1202 | 0.1245 | 0.1261 | 0.1026 | 0.1081 |
| D9 | 0.1440 | 0.1436 | 0.1493 | 0.1450 | 0.1625 | 0.1586 | 0.1535 | 0.1571 |
| D10 | 0.2300 | 0.2298 | 0.2517 | 0.2540 | 0.3046 | 0.3057 | 0.4769 | 0.4726 |
| MSE | 0.000001 | | 0.000005 | | 0.000009 | | 0.000024 | |
| MAE | 0.0005 | | 0.0018 | | 0.0026 | | 0.0043 | |
| MAS | 0.0016 | | 0.0043 | | 0.0063 | | 0.0098 | |
| IIM | 0.0001 | | 0.0004 | | 0.0011 | | 0.0049 | |
| K-S test | $D$-statistic = 0.1000 $p$-value = 1.000 | | $D$-statistic = 0.1000 $p$-value = 1.000 | | $D$-statistic = 0.1000 $p$-value = 1.000 | | $D$-statistic = 0.1000 $p$-value = 1.000 | |
| **Gini index** | **Observed** | **Estimate** | **Observed** | **Estimate** | **Observed** | **Estimate** | **Observed** | **Estimate** |
| | 0.287 | 0.287 | 0.315 | 0.316 | 0.411 | 0.411 | 0.590 | 0.589 |

**The error minimization method using all available grouped data on income shares by decile**. Besides the simple method, this study uses the specified parametric functional form that is constructed based on the weighted average of the exponential function and the functional form implied by Pareto distribution as shown in Eq. (1) in the "Methods" section and the curve fitting technique based on minimizing the sum of squared errors to estimate the Lorenz curve. Using all available grouped data on decile income shares of the same sample of countries from the UNU-WIID, the estimated parameters $P$ and $k$ are reported in Table 5. The values of $R^2$ ranging between 0.9995 and 1.0000 indicate that the estimated Lorenz curves fit the actual observations quite well. These results suggest that if more data on income distribution are available, the specified functional form that is based on the weighted average of the exponential function and the functional form implied by Pareto distribution could be used to directly estimate the Lorenz curve.

The estimated Lorenz curves using the error minimization method are then used to calculate the values of income shares by decile which would be compared to the actual observations. The results as reported in Table 6 show that the estimated decile income shares of four countries are practically close to the actual observations according to all goodness-of-fit statistical measures which are MSE, MAE, MAS, and IIM. The K-S test yields similar results in that the differences between the estimated decile income shares and the actual observations are not statistically significant with $p$-value equal to 1.000 in all cases.

Next, we use the income shares by decile estimated by using the error minimization method to calculate the value of the Gini index. The results also reported in Table 6 show that the estimated values of the Gini index of four countries are virtually similar to their actual observations as published in the UNU-WIID. Therefore, in case where more data on income distribution are available, the specified parametric functional form that is constructed based on the weighted average of the exponential function and the functional form implied by Pareto distribution could be used to directly estimate the Lorenz curve since it gives better results than using the simple method.

**An additional evaluation of the performance of the specified functional form for estimating the Lorenz curve that is constructed based on the weighted average of the exponential function and the functional form implied by Pareto distribution**. To further extend the evaluation of the performance of the specified functional form that is based on the weighted average of the exponential function and the functional form implied by Pareto distribution, we compare our alternative functional form to that proposed by Kakwani (1980). The reason that we choose Kakwani (1980)'s functional form is mainly because, according to

Cheong (2002) and Tanak et al. (2018), it is considered as the best performer among different functional forms for the Lorenz curve in fitting to the actual observations, most of which are often cited in the literature. Using the same notations for the cumulative normalized rank of income ($x$) and the cumulative normalized income ($y$) as denoted in the "Methods" section, the Kakwani (1980)'s functional form for estimating the Lorenz curve is as follows:

$$y(x) = x - a * x^{\alpha} * (1-x)^{\beta} \quad (8)$$

$$a > 0$$

$$0 < \alpha \leq 1$$

$$0 < \beta \leq 1$$

Cheong (2002) notes that, given the Kakwani (1980)'s functional form as shown in Eq. (8), the Gini index could be derived as $2a\ B(\alpha+1, \beta+1)$, where $B$ is the beta function.

Given the functional form proposed by Kakwani (1980) is found to be overall superior to other functional forms for the Lorenz curve as discussed above, this study employs the same dataset of four countries from the UNU-WIID and uses the Kakwani (1980)'s functional form along with the curve-fitting technique based on minimizing sum of squared errors to estimate the Lorenz curve. The estimated parameters $a$, $\alpha$, and $\beta$ as well as the values of $R^2$ are shown in Table 7. On the basis of the value of $R^2$, the Kakwani (1980)'s functional form performs slightly better than the alternative functional form reported in Table 5.

This study then uses the estimated Lorenz curves based on the Kakwani (1980)'s functional form to calculate the income shares by decile of four countries in the sample. We then compare the income shares by decile estimated using the Kakwani (1980)'s functional form to those estimated using the alternative functional form that is constructed based on the weighted average of the exponential function and the functional form implied by Pareto distribution. The results are reported in Table 8.

**Table 7 The estimated parameters $a$, $\alpha$, and $\beta$ for the Lorenz curves of four countries using Kakwani (1980)'s functional form.**

| Year | Country | $a$ | $\alpha$ | $\beta$ | $R^2$ |
|---|---|---|---|---|---|
| 2018 | Malta | 0.55 | 0.90 | 0.59 | 1.0000 |
| 2016 | Taiwan | 0.59 | 0.93 | 0.54 | 1.0000 |
| 2016 | USA | 0.78 | 0.96 | 0.53 | 1.0000 |
| 2015 | Cote d'Ivoire | 0.94 | 1.00 | 0.35 | 1.0000 |

The overall results illustrate that there are no significant differences among the values of income shares by decile estimated using the alternative functional form and the Kakwani (1980)'s functional form. Table 9 reports the comparison of the values of goodness-of-fit statistics between the alternative functional form and the Kakwani (1980)'s functional form.

The values of goodness-of-fit statistics as shown in Table 9 suggest that, besides the K-S test which gives identical results in that the differences between the estimated decile income shares and the actual observations are not statistically significant with $p$k-vkkalue equal to 1.000 in all cases, the performance of the Kakwani (1980)'s functional form is slightly better than that of the alternative functional form as measured by the values of MSE, MAE, MAS, and IIM.

However, when using both functional forms to calculate the Gini index, the estimated values of the Gini index computed using the alternative functional form are closer to the actual observations than those computed using Kakwani (1980)'s functional form. The comparison of the estimated values of the Gini index between the alternative functional form and the functional form proposed by Kakwani (1980) is reported in Table 10 which indicates that the alternative functional form that is constructed based on the weighted average of the exponential function and the functional form implied by Pareto distribution outperforms the Kakwani (1980)'s functional form. Note however that, for the alternative functional form, the value of the Gini index can be conveniently computed since it has a closed-form expression for the Gini index whereas, for the Kakwani (1980)'s functional form, the value of the Gini index is calculated by using the numerical integration since its explicit mathematical solution for the Gini index does not exist.

All in all, on the basis of the income shares by decile, the goodness-of-fit statistics, and the Gini index, it could be concluded that, given more data on income distribution are available, the overall performance of the alternative functional form is, by and large, comparable to that of the Kakwani (1980) which has the best overall performance among all functional forms according to Cheong (2002) and Tanak et al. (2018). However, our alternative functional form is more parsimonious than the functional form proposed by Kakwani (1980) since it has two parameters ($P$ and $k$) while Kakwani (1980)'s functional form has three parameters ($a$, $\alpha$, and $\beta$). The specified functional form used in this study is also computationally more convenient to calculate the Gini index than that of Kakwani (1980) which requires the valuation of the beta function or the numerical integration. In addition, the estimated values of the Gini index computed based on our specified functional form are closer to the values of the observed Gini index than those computed using the functional form proposed by Kakwani (1980). Furthermore, in the situation where the availability of data on income distribution is low, for example, only the bottom or the top or the ratio of the two, if we know the observed Gini index, our simple method that is developed based on the specified functional form could be used to approximate the Lorenz curve that would be closer the actual observations than that of Kakwani (1980) or any other functional forms that require more grouped data on income distribution than just 1 or 2 points.

**Table 8 The comparison of the estimated income shares by decile between the alternative functional form that is based on the weighted average of the exponential function and the functional form implied by Pareto distribution and the Kakwani (1980)'s functional form.**

| Decile | Malta | | | Taiwan | | | USA | | | Cote d'Ivoire | | |
|---|---|---|---|---|---|---|---|---|---|---|---|---|
| | Actual | Alternative | Kakwani (1980) | Actual | Alternaive | Kakwani (1980) | Actual | Alternative | Kakwani (1980) | Actual | Alternative | Kakwani (1980) |
| D1 | 0.0360 | 0.0376 | 0.0350 | 0.0336 | 0.0366 | 0.0345 | 0.0179 | 0.0242 | 0.0192 | 0.0074 | 0.0172 | 0.0091 |
| D2 | 0.0510 | 0.0500 | 0.0519 | 0.0491 | 0.0470 | 0.0484 | 0.0344 | 0.0320 | 0.0334 | 0.0178 | 0.0193 | 0.0168 |
| D3 | 0.0620 | 0.0612 | 0.0625 | 0.0590 | 0.0572 | 0.0583 | 0.0457 | 0.0423 | 0.0449 | 0.0263 | 0.0232 | 0.0251 |
| D4 | 0.0720 | 0.0720 | 0.0724 | 0.0684 | 0.0676 | 0.0680 | 0.0572 | 0.0545 | 0.0567 | 0.0353 | 0.0298 | 0.0346 |
| D5 | 0.0830 | 0.0829 | 0.0825 | 0.0779 | 0.0785 | 0.0782 | 0.0693 | 0.0686 | 0.0696 | 0.0459 | 0.0401 | 0.0458 |
| D6 | 0.0940 | 0.0943 | 0.0934 | 0.0890 | 0.0903 | 0.0895 | 0.0832 | 0.0847 | 0.0841 | 0.0583 | 0.0554 | 0.0595 |
| D7 | 0.1070 | 0.1068 | 0.1060 | 0.1022 | 0.1037 | 0.1029 | 0.1005 | 0.1033 | 0.1016 | 0.0759 | 0.0771 | 0.0774 |
| D8 | 0.1210 | 0.1219 | 0.1218 | 0.1199 | 0.1202 | 0.1202 | 0.1245 | 0.1261 | 0.1244 | 0.1026 | 0.1081 | 0.1033 |
| D9 | 0.1440 | 0.1436 | 0.1453 | 0.1493 | 0.1450 | 0.1469 | 0.1625 | 0.1586 | 0.1601 | 0.1535 | 0.1571 | 0.1497 |
| D10 | 0.2300 | 0.2298 | 0.2291 | 0.2517 | 0.2540 | 0.2529 | 0.3046 | 0.3057 | 0.3060 | 0.4769 | 0.4726 | 0.4787 |

**Table 9 The comparison of the values of goodness-of-fit statistical measures between the alternative functional form that is based on the weighted average of the exponential function and the functional form implied by Pareto distribution and the Kakwani (1980)'s functional form.**

| Goodness-of-fit measures | Malta | | Taiwan | | USA | | Cote d'Ivoire | |
|---|---|---|---|---|---|---|---|---|
| | Alternative | Kakwani (1980) | Alternative | Kakwani (1980) | Alternative | Kakwani (1980) | Alternative | Kakwani (1980) |
| MSE | 0.000001 | 0.000001 | 0.000005 | 0.000001 | 0.000009 | 0.000001 | 0.000024 | 0.000003 |
| MAE | 0.0005 | 0.0008 | 0.0018 | 0.0008 | 0.0026 | 0.0010 | 0.0043 | 0.0014 |
| MAS | 0.0016 | 0.0013 | 0.0043 | 0.0024 | 0.0063 | 0.0024 | 0.0098 | 0.0038 |
| IIM | 0.0001 | 0.00004 | 0.0004 | 0.0001 | 0.0011 | −0.0001 | 0.0049 | 0.0002 |
| K-S test | | | | | | | | |
| *D*-statistic | 0.1000 | 0.1000 | 0.1000 | 0.1000 | 0.1000 | 0.1000 | 0.1000 | 0.1000 |
| *P*-value | 1.000 | 1.000 | 1.000 | 1.000 | 1.000 | 1.000 | 1.000 | 1.000 |

**Table 10 The comparison of the estimated values of the Gini index between the alternative functional form that is based on the weighted average of the exponential function and the functional form implied by Pareto distribution and the Kakwani (1980)'s functional form.**

| Gini index | Malta | | | Taiwan | | | USA | | | Cote d'Ivoire | | |
|---|---|---|---|---|---|---|---|---|---|---|---|---|
| | Observed | Alternative | Kakwani (1980) | Observed | Alternative | Kakwani (1980) | Observed | Alternative | Kakwani (1980) | Observed | Alternative | Kakwani (1980) |
| | 0.287 | 0.287 | 0.281 | 0.315 | 0.316 | 0.308 | 0.411 | 0.411 | 0.401 | 0.590 | 0.589 | 0.569 |

## Conclusions

Finding a method for estimating the Lorenz curve is a practical and theoretical challenge. Many popular functional forms for the Lorenz curve do not have a closed-form expression for the Gini index, making it computationally inconvenient to calculate since they require the valuation of the beta function or the confluent hypergeometric function. In addition, despite the fact that the data on the observed Gini index are available and can be accessed from the well-known databases, such as the UNU-WIID, the World Bank's PovcalNet, and the WID, no study has utilized the observed Gini index in order to estimate the parameter(s) associated with the specified functional form. This is mainly due to the step that starts from using the reported Gini index and working backward to estimate the parameter(s) associated with the specified functional form is relatively more difficult compared with the conventional step which is to use the estimated parameter(s) associated with the specified functional forms to estimate the Gini index. To address these key issues, this study introduces a simple method for estimating the Lorenz curve with a closed-form expression for the Gini index. Our method utilizes three indicators, namely, the Gini index, the income share of the bottom, and that of the top for estimating the values of parameters associated with the specified functional form that is constructed based on the weighted average of the exponential function and the functional form implied by Pareto distribution without relying on any complicated error minimization technique, hence no software package is required for writing algorithms in order to estimate the parameters.

To demonstrate our simple method, the latest data on the Gini index and the income shares by decile and by percentile of four countries, namely, Malta, Taiwan, USA, and Côte d'Ivoire between 2015 and 2018 from the UNU-WIID are used. The overall results show that the estimated income shares by decile are practically close to their actual observations. This suggests that in case where the availability of data on income distribution is low, say, only 1 or 2 points, our simple method could still be used to retrieve the entire Lorenz curve, provided we know the observed Gini index, whereas other methods may need more grouped data on income in order to perform the same task. In addition to the simple method, we employ the specified parametric functional form and the curve-fitting technique based on minimizing sum of squared errors to estimate the Lorenz curve using all available grouped data on decile income shares. The estimated Lorenz curves for the same sample of countries indicate that our specified functional form satisfies the criteria for a good functional form as recommended by Dagum (1977) since it not only fits the actual observations quite well but also is not sensitive to different data on grouped income across countries that have a different level of income inequality, socioeconomic, and regional backgrounds. These estimated Lorenz curves are then used to compute the Gini index whose values are almost identical to their actual observations. Based on these results, our specified functional form could be used to directly estimate the Lorenz curve when more data on income distribution are available. Furthermore, using the same dataset of four countries, we show that our specified functional form has several advantages over the functional form proposed by Kakwani (1980), especially on the estimation of the Gini index.

Given the use of the Lorenz curve and the Gini index in various scientific disciplines (Eliazar and Sokolov, 2012), we hope that our simple method and the specified functional form for estimating the Lorenz curve could be applied to analyze size distributions of non-negative quantities and inequality.

## Data availability

The datasets generated and/or analyzed during the current study are included in this article and can be accessed from the United Nations University World Institute for Development Economics Research (UNU-WIDER) website (https://www4.wider.unu.edu/).

## Notes

1 See Eliazar and Sokolov (2012) for the applications of the Lorenz curve and the Gini index in other scientific disciplines.

2 See Sarabia (1997) and Ogwang and Rao (2000) for examples of parametric functional forms that combine two or more popular functional forms for the Lorenz curve using convex linear combinations. According to Ogwang and Rao (2000), a convex linear combination is a way to circumvent an important shortcoming of functional forms for the Lorenz curve which is the lack of satisfactory fit over the entire range of given income distribution.

3 It should be noted that given the observed Gini index is known, only the income share of the bottom $m$% or that of the top $m$% or the ratio of the two should suffice for estimating the entire Lorenz curve. This study uses three indicators because the Gini index gives the information about the general shape of the Lorenz curve, especially in the middle whereas the income shares of the bottom $m$% and the top $m$% provide the information at the tails of income distribution as discussed in the "Introduction" section.

## References

Aggarwal V (1984) On optimum aggregation of income distribution data. Sankhyā B 46:343–355
Arnold BC (1986) A class of hyperbolic Lorenz curves. Sankhyā B 48:427–436
Basmann RL, Hayes K, Slottje D, Johnson J (1990) A general functional form for approximating the Lorenz curve. J Econ 92:727–744
Bourguignon M, Saulo H, Fernandez RN (2016) A new Pareto-type distribution with applications in reliability and income data. Physica A 457:166–175. https://doi.org/10.1016/j.physa.2016.03.043
Chakrabarti BK, Chakraborti A, Chakravarty SR, Chatterjee A (2013) Econophysics of income and wealth distributions. Cambridge University Press, New York, pp. 1–6
Cheong KS (2002) An empirical comparison of alternative functional forms for the Lorenz curve. Appl Econ Lett 9:171–176. https://doi.org/10.1080/13504850110054058
Chotikapanich D (1993) A comparison of alternative functional forms for the Lorenz curve. Econ Lett 41:129–138. https://doi.org/10.1016/0165-1765(93)90186-G
Dagum C (1977) A new model of personal income distribution: Specification and estimation. In: Chotikapanich D (Ed) Modeling income distributions and Lorenz curves. Economic studies in equality, social exclusion and well-being, vol 5. Springer, New York, pp. 3–25
Eliazar II, Sokolov IM (2012) Measuring statistical evenness: a panoramic overview. Physica A 391:1323–1353. https://doi.org/10.1016/j.physa.2011.09.007
Fellman J (2018) Regression analyses of income inequality indices. Theor Econ Lett 8:1793–1802. https://doi.org/10.4236/tel.2018.810117
Gupta MR (1984) Functional form for estimating the Lorenz curve. Econometrica 52:1313–1314
Helene O (2010) Fitting Lorenz curves. Econ Lett 108:153–155
Jordá V, Sarabia JM, Jäntti M (2021) Inequality measurement with grouped data: parametric and non-parametric methods. J R Stat Soc Ser A 184:964–984. https://doi.org/10.1111/rssa.12702
Kakwani NC (1980) On a class of poverty measures. Econometrica 48:437–446
Kakwani NC, Podder N (1973) On the estimation of Lorenz curves from grouped observations. Int Econ Rev 14:278–292
Kakwani NC, Podder N (1976) Efficient estimation of the Lorenz curve and associated inequality measures from grouped observations. Econometrica 44:137–148
Lorenz MO (1905) Methods of measuring the concentration of wealth. Pub Am Stat Assoc 9:209–219
Ogwang T, Rao ULG (1996) A new functional form for approximating the Lorenz curve. Econ Lett 52:21–29
Ogwang T, Rao ULG (2000) Hybrid models of the Lorenz curve. Econ Lett 69:39–44
Ortega P, Martín G, Fernández A, Ladoux M, García A (1991) A new functional form for estimating Lorenz curves. Rev Income Wealth 37:447–452
Paul S, Shankar S (2020) An alternative single parameter functional form for Lorenz curve. Empir Econ 59:1393–1402. https://doi.org/10.1007/s00181-019-01715-3
Rao ULG, Tam AY-P (1987) An empirical study of selection and estimation of alternative models of the Lorenz curve. J Appl Stat 14:275–280. https://doi.org/10.1080/02664768700000032
Rasche RH, Gaffney JM, Koo AYC, Obst N (1980) Functional forms for estimating the Lorenz curve. Econometrica 48:1061–1062
Rohde N (2009) An alternative functional form for estimating the Lorenz curve. Econ Lett 105:61–63
Ryu H, Slottje D (1996) Two flexible functional forms for approximating the Lorenz curve. J Econom 72:251–274
Sarabia JM (1997) A hierarchy of Lorenz curves based on the generalized Tukey's lambda distribution. Econom Rev 16:305–320
Sarabia JM, Castillo E, Slottje D (1999) An ordered family of Lorenz curves. J Econ 91:43–60
Sarabia JM, Castillo E, Slottje D (2001) An exponential family of Lorenz curves. South Econ J 67:748–756
Sarabia JM, Pascual M (2002) A class of Lorenz curves based on linear exponential loss functions. Commun Stat–Theory Methods 31:925–942
Sarabia JM, Prieto F, Sarabia M (2010) Revisiting a functional form for the Lorenz curve. Econ Lett 107:249–252
Sarabia JM, Prieto F, Jordá V (2015) About the hyperbolic Lorenz curve. Econ Lett 136:42–45
Sarabia JM, Jordá V, Trueba C (2017) The Lamé class of Lorenz curves. Commun Stat–Theory Methods 46:5311–5326
Sitthiyot T, Budsaratragoon P, Holasut K (2020) A scaling perspective on the distribution of executive compensation. Physica A 543:123556. https://doi.org/10.1016/j.physa.2019.123556
Tanak AK, Mohtashami Borzadaran GR, Ahmadi J (2018) New functional forms of Lorenz curves by maximizing Tsallis entropy of income share function under the constraint on generalized Gini index. Physica A 511:280–288
Villaseñor JA, Arnold BC (1989) Elliptical Lorenz curves. J Econ 40:327–338
Wang Z, Smyth R (2015) A hybrid method for creating Lorenz curves. Econ Lett 133:59–63

## Acknowledgements

TS is grateful to Dr. Suradit Holasut for guidance and comments.

## Author contributions

Conceptualization: TS; Data curation: TS; Methodology: TS, KH; Formal analysis: TS; Validation: KH; Writing—original draft: TS; Writing—review and editing: TS, KH. Both authors read and approved the final manuscript.

## Competing interests

The authors declare no competing interests.

## Ethical approval

This article does not contain any studies with human participants performed by any of the authors.

## Informed consent

This article does not contain any studies with human participants performed by any of the authors.

## Additional information

**Correspondence** and requests for materials should be addressed to Thitithep Sitthiyot.